\documentclass[aps,prb,twocolumn,aps,floatfix,showpacs,showkeys,superscriptaddress]{revtex4-2}
\usepackage{amssymb}
\usepackage{amsbsy}
\usepackage{amsmath}
\usepackage{epsfig}
\usepackage{color}
\usepackage{float}
\usepackage{xr-hyper}
\usepackage{bm}
\usepackage[colorlinks,linktocpage,bookmarks=false,citecolor=blue,linkcolor=red,urlcolor=blue]{hyperref}

\newcommand{\beg}{\begin{gather}}
\newcommand{\eeg}{\end{gather}}
\newcommand{\beq}{\begin{equation}}
\newcommand{\eeq}{\end{equation}}
\newcommand{\bea}{\begin{eqnarray}}
\newcommand{\eea}{\end{eqnarray}}

\newcommand{\ep}{\epsilon}

\newcommand{\ket}[1]{\ensuremath{\left|#1\right\rangle}}

\newcommand\mean[1]{\ensuremath{\left\langle#1\right\rangle}}

\newcommand\lrp[1]{\left (#1\right)}
\newcommand\lrb[1]{\left[#1\right]}
\newcommand\lrc[1]{\left\{#1\right\}}
\newcommand\abs[1]{\left|#1\right|}

\newcommand{\lra}{\quad \Leftrightarrow \quad}

\newcommand{\be}{\begin{equation}}
\newcommand{\ee}{\end{equation}}
\def\ba{\begin{aligned}}
\def\ea{\end{aligned}}
\def\bes{\begin{subequations}}
\def\ees{\end{subequations}}
\def\bal{\begin{align}}
\def\eal{\end{align}}

\newcommand{\imk}[1]{{#1}}

\usepackage{soul}
\usepackage{ulem} 
\usepackage{hyperref}
\graphicspath{{./figures/}}

\begin{document}
\title{Re-entrant localization induced by short-range hopping in the fractal Rosenzweig-Porter Model }

\author{Roopayan Ghosh}
\thanks{ucaprgh@ucl.ac.uk}
\affiliation{Department of Physics and Astronomy, University College London, Gower Street, WC1E6BT, London}
\author{Madhumita Sarkar }
\affiliation{Department of Physics and Astronomy, University of Exeter, Stocker Road, Exeter EX4 4QL, United Kingdom}
\author{Ivan M. Khaymovich}
\affiliation{Nordita, Stockholm University and KTH Royal Institute of Technology Hannes Alfv{\'e}ns v{\"a}g 12, SE-106 91 Stockholm, Sweden}
\affiliation{Institute for Physics of Microstructures, Russian Academy of Sciences, 603950 Nizhny Novgorod, GSP-105, Russia}

\begin{abstract}
Typically, metallic systems localized under strong disorder exhibit a transition to \imk{delocalization} 
as kinetic terms increase. In this work, we reveal the opposite effect~--~increasing kinetic terms leads to an unexpected \imk{reduction of mobility, }
enhancing localization of the system, and even lead to re-entrant delocalization transitions. Specifically, we add a nearest-neighbor hopping with  amplitude \(\kappa\) to the Rosenzweig-Porter (RP) model with fractal on-site disorder and surprisingly see that, as \(\kappa\) grows, the system initially tends to localization from the fractal phase, but then re-enters the ergodic phase. We build an analytical framework to explain this re-entrant behavior, supported by exact diagonalization results. The interplay between the spatially local $\kappa$ term, insensitive to fractal disorder, and the energy-local RP coupling, sensitive to fine-level spacing structure, drives the observed re-entrant behavior. This mechanism offers a novel pathway to re-entrant localization phenomena in many-body quantum systems.

\end{abstract}

\maketitle


\paragraph{\textbf{Introduction:}} Ergodicity in quantum many-body systems~\cite{Deutsch1991,Srednicki1994} and its breakdown by a disordered potential~\cite{Basko06,gornyi2005interacting,Abanin_RMP}, usually called the many-body localization (MBL) transition has attracted a lot of attention for the past two decades. This stems both from fundamental challenges in determining whether the MBL phase persists in the thermodynamic limit~\cite{Luitz15,Sels2021,Vidmar19,DeRock16,Imbrie2016b,deroeck2024absencenormalheatconduction,sierant2024manybodylocalizationageclassical}, and from the potential applications of this phase in quantum information processing and machine learning~\cite{Smelyansky_Grover,Smelyansky_ML}.

According to common knowledge, disordered quantum systems typically exhibit localization behavior, such as in Anderson transition~\cite{Anderson1958} and many-body localization~\cite{Basko06,gornyi2005interacting}, which occur monotonically with increasing the disorder strength (or equivalently, decreasing the amplitude of the corresponding kinetic term). This behavior is believed to be universal both in many-body disordered systems~\cite{Abanin_RMP} and in the corresponding random-matrix (RM) ensembles utilized to understand them with Anderson~\cite{Evers2008Anderson} and ergodicity breaking~\cite{Kravtsov_NJP2015} phase transitions.
The latter models, like the so-called Rosenzweig-Porter random-matrix ensemble~\cite{Kravtsov_NJP2015,Biroli_RP,Ossipov_EPL2016_H+V, Monthus, BogomolnyRP2018,vonSoosten2017non,Venturelli2023replica,PhysRevResearch.5.043301}, show not only Anderson localized and ergodic (metallic) phases, but also an intermediate non-ergodic extended phase (relevant for the description of the Hilbert-space structure of the MBL states~\cite{QIsing_2021,Tarzia_2020,PhysRevLett.131.060404}), in which the eigenstates span over an extensive number but measure zero of all Hilbert-space configurations characterized by fractal dimensions $0\leq D_2\leq 1
$, see Eq.~\eqref{eq1}~\cite{Evers2008Anderson}.
\begin{figure}
    \centering
    \includegraphics[width=0.48 \linewidth,height=0.44\linewidth]{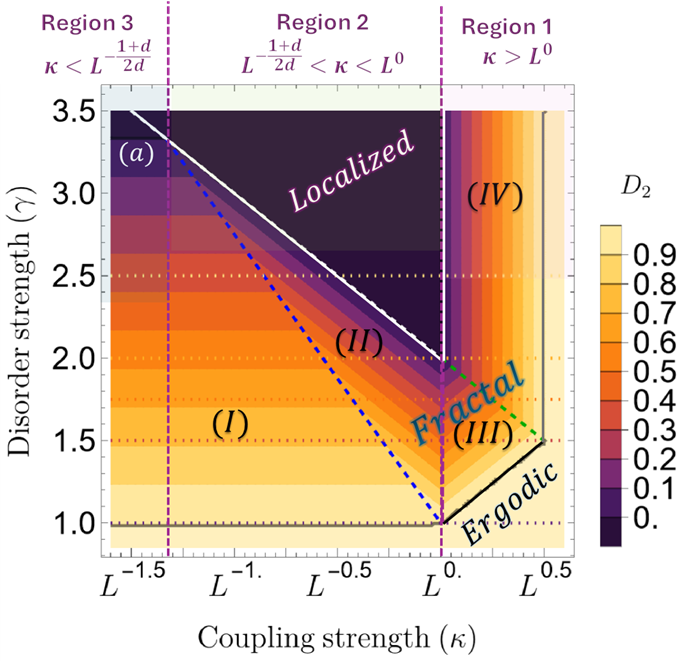}
    \includegraphics[width=0.45 \columnwidth]{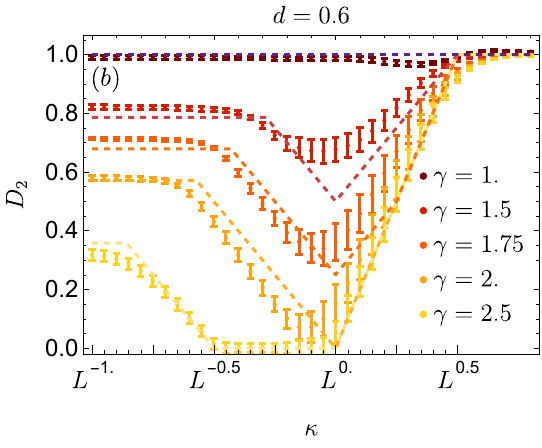}
  \caption{(a) Localization phase diagram in terms of fractal dimension $D_2$ for $d=0.6$ for different nearest neighbor coupling $\kappa$ and effective disorder strength ($\gamma$), obtained from exact analytical computation. \imk{The three regions described in text are demarcated by vertical dashed lines. Additionally,} the localized and fractal phases are separated by the white line and the fractal and ergodic phases are separated by the black line. Inside the fractal phase the regions with different behavior are separated by dashed lines and marked with roman numbers. The horizontal dotted lines denote the values of $\gamma$ chosen for (b).  (b)~$D_2$ with varying $\kappa$ for specific $\gamma$s. The numerical data shown via points are obtained by fitting IPR for different system sizes $2^n$ where $n$ runs from $7-14$ and averaged over $10^2-10^5$ realizations (the smaller system sizes have larger number of realizations). The dashed lines denote the analytical results.
  }
  \label{fig:schematic}
\end{figure}
In this letter, using RM ensembles, we analytically reveal a pathway that challenges the conventional view of a monotonic transition from localization to ergodicity. Starting with a system characterized by a fractal distribution of on-site disorder~\cite{Kravtsov2023Cantor, Sarkar2023_Fract_RP}, we show that adding a short-range kinetic term with coupling strength \(\kappa\) initially steers the system towards localization. Only beyond a critical strength of \(\kappa\), this trend reverses, and further increasing $\kappa
$ drives the system to an ergodic phase, see Fig.~\ref{fig:schematic}. Our analytical understanding provides a new perspective on re-entrant delocalization transitions \imk{in the disordered systems, robust to perturbations}, a phenomenon not previously explored in the literature~\imk{\footnote{\imk{Here we don't consider the models with deterministic quasiperiodic potentials as they are crucially sensitive to random perturbations and immediately undergo the localization transition for any random perturbation}.}}.
We then verify our predictions with exact numerical simulations.


\paragraph{\textbf{Model:}}
The standard Rosenzweig-Porter model~\cite{RP,Kravtsov_NJP2015} is a random-matrix model where both the diagonal and off-diagonal elements of the $L \times L$ Hermitian matrix $H_{mn} = H_{nm}^*$ are taken from a random normal distribution such that,
\begin{equation}
    H_{mn}=h_n \delta_{mn}+R_{mn}\cdot L^{-\gamma/2} \ .
    \label{eq:RP}
\end{equation}
Here $\overline{h_n}=\overline{R_{mn}}=0$ and $\overline{h_n^2}=\overline{\imk{\abs{R_{mn}}}^2}=1
$ and $\gamma
$ is the tuning parameter which rescales the off-diagonal elements and controls the phase diagram. The ergodic (\(\gamma \leq 1\)) and localized (\(\gamma \geq 2\)) phases are separated by a non-ergodic extended fractal phase (\(1 \leq \gamma \leq 2\)), which is squeezed between the ergodic transition (ET) at \(\gamma = 1\) and the Anderson localization transition (AT) at \(\gamma = 2\). The eigenstate nature in different phases can be characterized by the  fractal dimension $D_2$, calculated from the eigenfunctions $\ket{\psi}
$, as~\cite{Evers2008Anderson}
\begin{equation}
  L^{-D_2}= {\rm IPR}= \sum_{i=1}^L|\langle i | \psi \rangle|^{4} \label{eq1}
\end{equation}
where $\ket{i}$ denotes the computational basis states and ${\rm IPR}$ denotes the inverse participation ratio. \imk{When we compute $D_2$ numerically, we fit the IPR for different system sizes $L$ and extrapolate to obtain the thermodynamic limit of $D_2$. Corrections $O(1/\log L)$ to $D_2$ are known to be present for the RP models~\cite{10.21468/SciPostPhys.18.3.090}, which were also seen in our case and were taken into account during fitting.} The ergodic states are characterized by $D_2=1$, the extended (multi)fractal states are characterized by $0<D_2<1$ and for the localized state $D_2=0$. Recently it has been shown~\cite{Sarkar2023_Fract_RP} that the fractal distribution of diagonal elements $h_n$'s tunes the phase diagram of the Rosenzweig-Porter (RP) model by extending the non-ergodic fractal phase. Simply put, in this scenario \imk{of fractal diagonal disorder} one considers \imk{the level spacing} $s_i=h_{n_{i+1}}-h_{n_i}$\imk{, where $h_{n_i}<h_{n_{i+1}}$ are sorted diagonal elements $h_n$,to be i.i.d. numbers, } distributed as a Pareto distribution $P (s) = \frac{d\cdot \delta_{typ}^d}{s^{d+1}}\theta (s-\delta_{typ})$
where $\delta_{typ}\sim L^{-1/d}$ is the typical level spacing. \imk{The elements $h_n$ are randomly shuffled with respect to the sorted sequence $h_{n_i}<h_{n_{i+1}}$ to realize random disorder.}
Consequently, the number of elements $h_{n}$ in the interval $L^{-b}$ around a certain energy $E$ scales as $N\{\abs{E-h_n}\in [L^{-b-db},L^{-b}]\}\sim L^{1-f(b)}db$~\imk{\footnote{\imk{Here and further, we neglect all the $L$-dependencies slower than any power and look at the model in terms of the saddle-point approximation.}}}, where $f(x)=d\cdot x, df(x)/dx\leq 1$ for Hermitian matrices, $f(0)=0$ and $d
$ can be regarded as the fractal dimension of the disorder. In this work, we add an extra nearest-neighbor hopping term to modify Eq.~\eqref{eq:RP} to:

\begin{equation}
    H_{mn} = h_n \delta_{mn} + \kappa(\delta_{m,n+ 1}+h.c.)  + R_{mn} L^{-\gamma/2} \ ,
    \label{eq:nnfRP}
\end{equation}
where \( h_n \) is distributed as the aforementioned fractal disorder~\cite{SM}, and the amplitude of the extra nearest-neighbor hopping term scales with the system size, \( \kappa \sim L^k \), with positive or negative $k$. \imk{In what follows, we} shall \imk{focus on the properties of the bulk spectral states and} discuss the non-monotonic \imk{behavior of their fractal dimension} by tuning this extra term $k$.

\paragraph{\textbf{Phase diagram:}}
By increasing the nearest-neighbor hopping strength \( \kappa \) from zero, a strikingly counterintuitive phenomenon occurs. For small but significant coupling, the fractal dimension \( D_2 \) initially decreases, driving the system towards localization. Only beyond a certain critical hopping amplitude  \( \kappa \) does the standard increasing behavior of \( D_2 \) (indicating more delocalization) dominates. This reveals a rich re-entrant phase diagram as shown in Fig.~\ref{fig:schematic}(a), where we plot the fractal dimension ($D_2$) vs varying coupling strength ($\kappa$) and disorder strength (which is controlled by $\gamma$), see~\cite{SM} for results for other values of $d$. From a fully analytic study described later, we compute the fractal dimension $D_2$ of the different regions of the diagram as follows:

\subsection{Region 1}
For $k>0$, $\kappa\gg 1$ there are no effects of the fractal diagonal disorder:
\be
D_2(\gamma) =
\left\{
  \begin{array}{ll}
    1, & \phantom{\gamma_{ET}^{(1)}<}\gamma<\gamma_{ET}^{(1)} 
    ,\\
    2-\gamma+k, & \gamma_{ET}^{(1)}<\gamma<\gamma_P, \\
    2k, & \gamma_{P\phantom{T}}<\gamma. 
  \end{array}
\right. \ ,
\label{eq:region1}
\ee
because of the spatial locality of the dominant hopping term $\kappa$. Here, there is no localized phase, only the ergodic, $\gamma<\gamma_{ET}^{(1)}=1+k
$, and fractal phases. We distinguish the two fractal phases as RP-like  (\textbf{fractal III} in Fig.~\ref{fig:schematic}(a)) for the middle regime, and block-like (\textbf{fractal IV}), $\gamma>\gamma_P=2-k$, for the final regime in Fig.~\ref{fig:schematic}(a). This is because  fractal IV has energy level spacings distributed according to Poisson statistics in spite of it not being localized, distinct from the other regime. 
The middle regime disappears at $k=1/2$ and the system becomes ergodic for all $k>1/2$ at any $\gamma
$.

\subsection{Region 2}
For intermediate $-(1+d)/(2d)<k<0$, there is competition between the 
hopping terms and the fractal disorder resulting in four distinct regimes of $\gamma$:
\be
D_2(\gamma) =
\left\{
  \begin{array}{ll}
    1, & \phantom{\gamma_{AT\;}<}\gamma<\gamma_{ET}^{(2)} 
    ,\\
    \frac{2-d\cdot\gamma}{2-d}, & \gamma_{ET}^{(2)}<\gamma<\gamma_{FT} 
    ,\\
    2-\gamma+|2k|\frac{2-2d}{1+d}, & 
    \gamma_{FT}<\gamma<\gamma_{AT}^{(2)} 
    ,\\
    0, & \gamma_{AT}^{(2)}<\gamma 
  \end{array}
\right.
\label{eq:region2}
\ee
 ergodic, $\gamma<\imk{\gamma_{ET}^{(2)}}=1
 $ and localized, $\gamma>\gamma_{AT}^{(2)}= 2+\frac{2-2d}{1+d}|2k|$ ones are complimented by two fractal phases: the fractal-disorder-dominated~\cite{Sarkar2023_Fract_RP}, $\gamma<\gamma_{FT} = 1+\frac{2-d}{1+d}|2k|$ (\textbf{fractal I} in Fig.~\ref{fig:schematic}(a)) and the standard RP-like~\cite{Kravtsov_NJP2015}, $\gamma>\gamma_{FT}$ denoted by \textbf{fractal II} in Fig.~\ref{fig:schematic}(a) to distinguish their origins.

\subsection{Region 3}
For small $k<-(1+d)/(2d)
$, there are no effects of the short-range hopping and the results coincide with the ones of~\cite{Sarkar2023_Fract_RP} with the enlarged fractal-disorder-dominated phase, also denoted by \textbf{fractal I}:
\be
D_2(\gamma) =
\left\{
  \begin{array}{ll}
    1, & \phantom{\gamma_{ET}<}\gamma<\gamma_{ET}^{(3)}
    ,\\
    \frac{2-d\cdot\gamma}{2-d}, & \gamma_{ET}^{(3)}<\gamma<\gamma_{AT}^{(3)},\\
    0, & \gamma_{AT}^{(3)}<\gamma
    .
  \end{array}
\right.
\label{eq:region3}
\ee
with $\gamma_{ET}^{(3)} = 1$ and $\gamma_{AT}^{(3)} = 2/d$.
The re-entrant behavior is even more evident in Fig.~\ref{fig:schematic}(b) by considering fixed \(\gamma\) cuts from Fig.~\ref{fig:schematic}(a), marked by the dotted lines on the plot.

The numerical data, represented by points, aligns well with the analytical prediction of the phase diagram outlined above, which is plotted alongside the data with dashed lines. As is clearly evident for $\gamma>1$ $D_2$ shows a reduction with increasing value of $\kappa$ first, before reverting back to the standard behavior towards delocalization.

In what follows, we provide a detailed analytical explanation for the different regions of the phase diagram. In summary, the key insight lies in identifying how the nearest-neighbor coupling hybridizes the fractal RP eigenstates, thereby modifying the spectral distribution. In certain parameter regimes, due to the hybridization, the effective on-site disorder distribution is transformed from correlated fractal to uncorrelated Gaussian, which triggers localization. We shall treat the cases \(k > 0\) and \(k < 0\) separately, as they give rise to distinct phenomena.

\paragraph{\textbf{Analytical explanation:}}
Let us first recall the main results for $\kappa=0
$ i.e. $k \rightarrow -\infty$, the fractal RP model limit.
The eigenfunction of any RP model has a Lorentzian structure~\cite{Monthus,BogomolnyRP2018,2022_nonHerm_RP} (including the fractal disorder case~\cite{Sarkar2023_Fract_RP})
\be\label{eq:psi_RP_Lorenzian}
|\psi_m(n)|^2 = \frac{A}{\lrp{h_n - E_m}^2 + \Gamma_d^2}
\ee
where $A$ is the normalization coefficient. We also have $\Gamma_d \sim L^{-a}$ with the parameter $a$ being the solution of the self-consistency equation
\begin{equation}
1 + 2a - f(a)=\gamma.
\label{eq:selfconsistency}
\end{equation}
Clearly, $\Gamma_d$ is the width of the eigenfunction miniband, within which all the eigenstates are fully hybridized, and is found to be~\cite{Sarkar2023_Fract_RP}.
\begin{equation}
    \Gamma_d\sim L^{-\frac{\gamma-1}{2-d}}
    \label{eq:gammad}
\end{equation}
For small enough short-range hopping $\kappa
$, the fractal phase is dominated by the local-in-energy RP broadening $\Gamma_d$~\cite{Kravtsov_NJP2015}, highly sensitive to the fractal on-site disorder~\cite{Sarkar2023_Fract_RP}.
At large $\kappa\gtrsim 1$, the standard picture of $1D$ Anderson prevails~\cite{Das2023beta-ALT}, removing all the fine (fractal) details of the diagonal-energy distribution via strong short-range resonances.
In the intermediate region the competition between $\Gamma_d$ and $\kappa$ comes into play and gradually destroys the fractal structure of the diagonal disorder: the interplay between two delocalizing mechanisms reduces the fractal dimension and can even localize the system.
Let us now consider the situation in details for the two different limits,
\\

\paragraph{\textbf{ Strong nearest-neighbour hopping $\mathbf{k>0 \implies \kappa \gg1}$ (Region 1 ): }}
\begin{figure}
    \centering
    \includegraphics[width=0.98 \columnwidth]{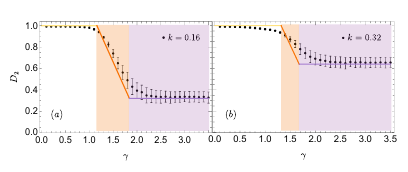}
    \caption{Comparison of numerically computed variation fractal dimension $D_2$ with $\gamma$ (points with error bars) to analytically computed Eq.~\eqref{eq:region1} (solid lines) for $k=0.16$ and $k=0.32$. The orange (lilac) line and shaded regions denote fractal III (IV) phase, yellow line stands for the ergodic phase. 
    }
    \label{fig:kgrt0}
\end{figure}
When $\kappa \gg 1$, the nearest-neighbor term dominates the Hamiltonian and strongly hybridizes nearby sites, effectively creating a block diagonal structure in the Hamiltonian, before the effects of RP-coupling kick in. The block sizes scale as $
\xi\sim \kappa^2= L^{2k}\gg 1
$, see~\cite{Barney2023blockRP,Das2023beta-ALT}, where $\xi$ is the localization length in an effective $1$D Anderson model with disorder strength $\kappa^{-1}$~\cite{Izrailev1998ALT,Das2023beta-ALT}. Within each block, wave functions become fully hybridized, rendering the fractal disorder distribution in diagonal elements $h_n$ insignificant.
The bandwidth of the $\xi \times \xi$ block is equal to the total bandwidth and can be estimated as, $E_{BW}^2 =\frac{1}{L} \sum_{m,n=1}^{L} \mean{\abs{H_{m n}}^2} =\frac{1}{\xi} \sum_{m,n=1}^{\xi} \mean{\abs{H_{m n}}^2} = 2\kappa^2 + 1
$ .
The corresponding mean level spacing $\delta_\xi$ in each of such blocks reads as $\delta_\xi \simeq \frac{E_{BW}}{\xi} $ ,
as the number of levels in each block is given by the block size $\xi$.

Using $\kappa\sim\xi^{1/2}\gg1$, we have $E_{BW} \simeq \xi^{1/2}$. Therefore the local level spacing in each block, $\delta_\xi\sim \xi^{-1/2}$ is parametrically larger than the global one $\delta_L = E_{BW}/L\simeq \xi^{1/2}/L$ as soon as $\xi\ll L$.
For $\xi\gtrsim L$ there is only one block, therefore $\delta_{\xi} = \delta_L$ and all the wave functions are ergodic $D_2=1$. This happens when $k >1/2$~\cite{Das2023beta-ALT}.
Therefore in what follows, we focus on the case $0\leq k<1/2$.

For such $k$,
the effect of fractal distribution is suppressed due to the strong hybridization in each block\imk{, (also see the consideration of region 2 below with $b_*<0$ for the current case of $k>0$)} and hence the RP all-to-all coupling will hybridize all the energy levels of the blocks in the interval $\Gamma_1$, similar to the case of non-fractal disorder $d=1$~\cite{Kravtsov_NJP2015}, where $\Gamma_1$ is the width of the Lorentzian eigenfunction with magnitude, $\Gamma_1 \sim L^{1-\gamma}$, which replaces $\Gamma_d$ in Eq.~\eqref{eq:psi_RP_Lorenzian},
according to Fermi's Golden rule. In order to find the wave-function fractal dimension, we should now compare $\Gamma_1$ to the block-local $\delta_\xi$ and global $\delta_L$.

For \bm{$\Gamma_1<\delta_L\ll\delta_\xi$} we have no additional hybridization given by $\Gamma_1$ and the fractal support set $L^{D_2}$ is given by
$D_2 = \frac{\ln \xi}{\ln L} = 2k$
similar to a generalized $1$D Anderson model~\cite{Das2023beta-ALT}. This corresponds to the limit $\gamma>2-k$ and constitutes the fractal IV behavior with Poisson level statistics.
For \bm{$\delta_L<\Gamma_1<\delta_\xi$}, in the energy interval $\Gamma_1$ there is at most $O(1)$ levels from each block. Therefore the number of \imk{blocks, hybridized by $\Gamma_1$, will be equal to the number of hybridized levels} $\Gamma_1/\delta_L$. As a result the fractal support sets $\sim \xi$ of all $\Gamma_1/\delta_L$ hybridized wave functions, combined together, will determine the fractal dimension as (cf.~\cite{Barney2023blockRP})
\be\label{eq:D_2_blockRP_0}
L^{D_2} = \xi\cdot \frac{\Gamma_1}{\delta_L}\lra D_2 = 2-\gamma + k \ ,
\ee
where we used the fact that for $\xi\gg 1$, $\delta_\xi = \xi^{-1/2}\sim L^{-k}$ and that the support set of each eigenstate in each block is $\xi$. This behavior constitutes the fractal III RP-like region.
For \bm{$\Gamma_1 > \delta_{\xi}$},i.e., $\gamma<1+k$, the fractal dimension~\eqref{eq:D_2_blockRP_0} saturates at $D_2 = 1$
and the wave functions are ergodic. Therefore for $k>0$, we reproduce Eq.~\eqref{eq:region1}.

There are several things to note here. Firstly, since $k>0
$, in the fractal regime, if present, the fractal dimension of the eigenfunctions are bigger than in the usual RP case~\cite{Kravtsov_NJP2015}, $D_2>2-\gamma$. Secondly, the ergodic transition now shifts to $\gamma_{ET}^{(1)} = 1+k$ and there is no localization transition $\gamma_{AT}^{(1)}$ for any $\gamma$, unlike the standard RP models. What we instead have is a change of level statistics to  Poisson statistics at $\gamma_P = 2-k$, while the wave functions are extended with a $\gamma$-independent fractal dimension of $2k$ within each of nearly independent blocks of size $\xi$~\cite{Das2023beta-ALT,Tang2022nonergodic}. This is a hallmark of the extensively varying local hopping term.
At $k=1/2$ the ergodic $\gamma_{ET}$, and Poisson $\gamma_{P}$, transitions merge simultaneously with the change of the statistics at $\gamma>\gamma_P$ from Poisson to Wigner-Dyson and, thus, only ergodic phase survives. \imk{The above predictions for the fractal dimensions, Eq.~\eqref{eq:D_2_blockRP_0}} are verified with exact numerics in Fig.~\ref{fig:kgrt0}, where the solid lines represent Eq.~\eqref{eq:region1}.

\paragraph{\textbf{Weak nearest-neighbor hopping, $\mathbf{k \leq0 \implies \kappa\lesssim1}$ (Regions 2 and 3)}:}
\begin{figure}
    \centering
     \includegraphics[width=0.98 \columnwidth]{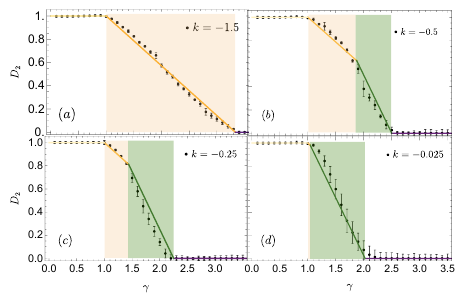}

    \caption{(a)$D_2$ vs $\gamma$ for $k=-1.5$ which is in Region 3. The solid line is Eq.~\eqref{eq:region3}. (b),(c),(d): Three values of $k$ showing the behavior in Region 2. The solid lines are obtained from Eq.~\eqref{eq:region2}, yellow, dark yellow, dark green and purple denote ergodic, fractal I, fractal II and localized phases respectively. Fractal I and II regions are also shaded by yellow and green colours respectively.
    }
    \label{fig:kless0}
\end{figure}
In this region, we first need to recompute the localization length \(\xi\) and the hybridized wave functions. This step is crucial, as our primary goal is to accurately compute how hybridization due to the additional term impacts the fractal-distributed disorder \(h_n\). However, since \(\kappa\) is small, we can employ perturbation theory~\cite{Izrailev1998ALT}.

If the wave function \(\psi_m(n)\) reaches its maximum at \(n = m\), then, applying perturbation theory for \(\kappa < 1\), we obtain $
\psi_m(m\pm1) =\frac{\kappa}{h_m-h_{m\pm1}} \psi_m(m)\simeq \psi_m(m) e^{-1/\xi} \ ,
$
from which we can extract
$
\xi^{-1} = -\ln \kappa + \mean{\ln|h_m|} = -\ln\lrp{2 e \kappa}
$
by approximating $h_n$s via a box distribution with $|h_m|<1/2$~\cite{Izrailev1998ALT}.
The corresponding energy shift is given by the second order perturbation theory
\be\label{eq:hn'-hn}
\epsilon_m = h_m +\sum_{\eta=\pm1}\frac{\kappa^2}{h_m - h_{m+\eta}} \ ,
\ee
which typically shifts the diagonal disorder, $h_m$, by
\be\label{eq:ep_n_shift_by_kappa<<1}
\exp\lrp{\mean{\ln|\epsilon_m - h_m|}} \simeq \kappa^2 \ .
\ee
Beyond the typical shift, the probability to have $\abs{\epsilon_m - h_m}\sim L^{-b}\imk{\gg} \kappa^2$, i.e., with $b<-2k$, is given by the probability $p_b$ to have small
$\abs{h_m - h_{m\pm 1}}\sim L^{b+2k}\ll 1$ ~\footnote{Note that $b>-2k>0$ is not achievable, because $|h_m-h_{m\pm1}|\leq O(1)$.}.
\imk{Note that, as before here we consider the randomly reshuffled diagonal elements $h_n$, but not the monotonic sequence $h_{n_i}<h_{n_{i+1}}=h_{n_i}+s_{i}$, given by i.i.d. $s_i>0$. Therefore the marginal distribution $P(h_m - h_{m\pm 1})$ of $h_m - h_{m\pm 1}$ is regular close to the origin.
For such distributions, regular at the origin,} $p_b$ is proportional to the small interval, $\abs{h_m - h_{m\pm 1}}$, i.e. $ p_b\sim L^{b+2k}\ll 1$.
Thus, among $L$ differences \imk{$\abs{h_m - h_{m-1}}$} in the entire sample there are
$M_b \sim L\cdot p_b\sim L^{b+2k+1}$ levels shifted by $\delta h \equiv \abs{\epsilon_m - h_m}\sim L^{-b}$.
Note that as we have only $L$ level differences $\delta h$,
the typical maximal energy hybridization interval $\delta h_{\max}$, reaches only at $M_b\sim L^0$ levels, and scales as
\imk{$\delta h_{\max}\sim L^{2k+1}$}.
Thus, $\delta h_{\max}$ will play the role of the crucial energy scale in this regime.
\\

\paragraph{Region 2:}
\imk{In the} case when \bm{\imk{$\delta h_{\max}\gg \delta_{typ}$}}, \imk{ which we denote by} Region 2, the hybridization from the \imk{short-range} hopping \imk{already} affects the fractal diagonal disorder \imk{($\delta h_{\max}\gg \Gamma_d$) at least at some $\gamma$ values}, but the corresponding block RP picture is not applicable as the block size $\xi_\kappa\simeq \lrb{\ln\lrp{1/(2 e \kappa)}}^{-1}\ll1$ is small~\cite{Izrailev1998ALT}. 

\imk{As $\xi\ll 1$, the above effect of short-range hopping  is significant only when the hybridization in energy larger than the typical level spacing $\delta_{typ}\sim L^{-1/d}$.
This leads to the following condition
\be\label{eq:dh_max_vs_delta_typ}
\delta h_{\max}\gg \delta_{typ} \lra -1-1/d<2k<0 \ .
\ee

Now let's focus on the effect of the nearest-neighbor hopping on the fractal diagonal disorder. For this, we will compare the fractal structure of the latter with the hybridization $\delta h$, shifting this levels. Indeed, t}he number of fractal levels in the interval $L^{-b}
$ are given by $L^{1-f(b)}$, while the number of levels shifted by the same energy $\delta h\sim L^{-b}$ is $M_b\sim L^{1+b+2k}$.
This means that as soon as $L^{1-f(b)}\ll M_b$, corresponding to \ $b>b_*$, all the levels in the interval $L^{-b}$ will be hybridized by at least $\delta h \sim L^{-b}$ and, thus, losing their fractal distribution they will be redistributed homogeneously on that interval $L^{-b}$.
Otherwise, if $L^{1-f(b)}\gg M_b$, there will just be a measure zero of such hybridization events with a negligible impact on counting.

For small $b < b_*
$ (large enough intervals $L^{-b}$), we have $L^{1-f(b)} > M_b$. Since $f(b) = d \cdot b$ at such $b<b_*$, the solution of the equation $L^{1-f(b_*)} = M_{b_*}$ is given by
\be\label{eq:b*}
1 - |2k| + b_* = 1 - d \cdot b_* \quad
\Longrightarrow \quad b_* = \frac{|2k|}{1 + d},
\ee
which describes the intervals $L^{-b_*}$, smaller than which everything in the interval is fully hybridized, i.e., $df(b)/db = 1$.
Consequently, the corresponding $f(b)$, changed by $\kappa$-elements, can be read from the continuity equation at $b=b_*$ 
as
\be
\label{eq:newfb}
f(b) =
\left\{
  \begin{array}{ll}
    d\cdot b, & b<b_* \\
    b-(1-d)b_*, & b_*<b<b_{**}
  \end{array}
\right. \ ,
\ee
where $b_{**}=1+(1-d)b_*$ is determined by $f(b_{**})=1$.

This means that as soon as the solution of the self-consistency Eq.~\eqref{eq:selfconsistency} corresponds to $f(b<b_*)=d\cdot b$, i.e., at $\gamma<\gamma_{FT} = 1+(2-d)b_*$, the fractal I solution~\cite{Sarkar2023_Fract_RP} is valid,
\imk{given by the second line in Eq.~\eqref{eq:region2} using $\Gamma_d$ from~\eqref{eq:gammad}.}
In the opposite case of $\gamma>\gamma_{FT}$, substituting $f(b)$ from Eq.~\eqref{eq:newfb} to Eq.~\eqref{eq:selfconsistency} gives
\be\label{eq:D_negative_k}
\Gamma_\kappa = L^{b_{**}-\gamma}, \quad D_2(\gamma>\gamma_{FT}) = 2b_{**}-\gamma
\ee
It works until the Anderson transition at $\gamma_{AT}^{(2)} = 2 b_{**}=2+\frac{2-2d}{1+d}|2k|$, 
\imk{describes fractal II region, matching all the other limiting cases, and corresponds to the third line in Eq.~\eqref{eq:region2}.}

To corroborate the analytical results we plot numerical data for $D_2$ vs $\gamma$ for $k=-0.5,-0.25,-0.025$ in Fig.~\ref{fig:kless0}(b)-(d) with the analytical predictions shown in solid lines. Here, dark yellow and dark green solid lines have been used to denoted fractal I and II behavior, respectively. The numerical data matches very well with our analytical model.
\\

\paragraph{Region 3:}
\imk{This is the scenario where the entire phase diagram remains unaffected by short-range hopping and is exactly predicted by the analysis in Ref.~\onlinecite{Sarkar2023_Fract_RP}. From Eq.~\eqref{eq:newfb}, one can see that this occurs when \( b_{**} \le b_* \), effectively eliminating any trace of the fractal II region from the phase diagram. This yields the condition:
\begin{equation}
    1 + \frac{1 - d}{1 + d}|2k| \le \frac{|2k|}{1 + d} \quad \Leftrightarrow \quad |2k| \ge 1 + \frac{1}{d}
\end{equation}

This is also fully consistent with the condition \( \delta h_{\max} \ll \delta_{\mathrm{typ}} \), which is the reverse of the inequality provided in Eq.~\eqref{eq:dh_max_vs_delta_typ}. In this case, the maximal energy hybridization interval \( \delta h_{\max} \) of the generalized model is smaller than \( \Gamma_\kappa \) for all values of \( \gamma \), which in turn upper bounds \( b_* \).

Although for \( \Gamma_k \gg \delta_{\mathrm{typ}} \) the above condition may initially appear to be stronger than necessary, it is important to ensure that this inequality holds throughout the fractal regime—including regions near localization where \( \Gamma_k \ll \delta_{\mathrm{typ}} \).
}
\imk{This describes Eq.~\eqref{eq:region3}, which is given by results of Ref.~\onlinecite{Sarkar2023_Fract_RP}.}
In Fig.~\ref{fig:kless0}(a) we numerically compute the phase diagram of fractal dimension $D_2$ for $d=0.6$ and \imk{$k=-1.5$, i.e. $|k|>1/2+1/(2d)$} which shows excellent agreement with our theoretical prediction.

\paragraph{\textbf{Conclusion:}}
In this letter, utilizing the Rosenzweig Porter model with fractal disorder, we show that a short-range kinetic term may not always drive the system towards delocalization. In fact, we analytically demonstrate that increasing amplitude of this term results in an unexpected non-monotonic behavior of the localization phase diagram and leads to the re-entrant delocalization. This work elevates the tunability of the fractal phase in RP model, as introduced in~\cite{Sarkar2023_Fract_RP}, to a new level. It shows that by applying local perturbations, a system with fractal diagonal disorder can be driven to localized, fractal, or ergodic states.
The underlying mechanism is based on the interplay between the local-in-space tight-binding hopping term $\kappa$ and the local-in-energy Fermi's golden rule broadening $\Gamma_d$ in the Rosenzweig-Porter model. In the intermediate regime, these competing delocalization mechanisms reduce the fractal dimension and can even induce localization.
In quantum many-body systems, locality emerges in energy, real, and Hilbert space. Extending this mechanism from random matrix models to such systems could enable competition between localization in different spaces, potentially inducing re-entrant ergodic transitions in many-body setting.

\begin{acknowledgments}
\paragraph{\textit{Acknowledgements:}}R.~G. and M.~S. thanks Lev Vidmar for discussions. I.~M.~K. acknowledges the support
by the European Research Council under the European
Union's Seventh Framework Program Synergy
ERC-2018-SyG HERO-810451. R.~G. acknowledges support by EPSRC grant MACON-QC EP/Y004590/1.
M.~S. acknowledges the support from UK EPSRC award under the Agreement EP/Y005090/1.
\end{acknowledgments}
\bibliography{Lib}

\appendix
\section{Details of Fractal RP model}
\label{app:appA}
Let us provide a brief description of the fractal RP model. In this special RP model, we consider the diagonal elements ($h_n$) to be chosen from certain (multi)fractal distribution (e.g., on a Cantor set~\cite{Kravtsov2023Cantor,Sarkar2023_Fract_RP}).
In such a case, one can just count how many $n$s are distributed in the interval
$|E-h_n|\in \lrb{L^{-b-db},L^{-b}}$, parameterized by $b \ge 0$,
\be\label{eq:f(b)_def}
L^{1-f(b)}\equiv \#\lrc{n:|E-h_n|\in \lrb{L^{-b-db},L^{-b}}} \ ,
\ee
with a certain $f(b)$, characterizing the above multifractal.
We assume that the overall bandwidth of the diagonal elements is $O(1)= O(L^0)$, thus, in general $f(0)=0$.

For example, if we choose $h_n$s from a Cantor set,
\be\label{eq:f(b)_Cantor}
f(b) = d \cdot b \ ,
\ee
with $d$ being a Hausdorff dimension of the Cantor set.
The usual uncorrelated random case~\cite{Kravtsov_NJP2015} corresponds to $d=1$, while the non-Hermitian complex one~\cite{2022_nonHerm_RP} gives $d=2$.
In principle, one can consider any values of $d$ in the interval $0\leq d\leq 2$ if one considers the non-Hermitian matrices with complex entries~\cite{2022_nonHerm_RP}.

In general, $f(b)$ has the following properties
\begin{itemize}
  \item $f(0)=0$ and $df(b)/db \geq 0$ as we assume that all $L$ diagonal elements are within the bandwidth $L^0$ and their number decays with the decaying interval $|E-\ep_n|\sim L^{-b}$;
  \item $f(\Delta_{typ}) = 1$ with $\delta_{typ} \sim L^{-\Delta_{typ}}$ being the typical level spacing of the set. Note that in the real case the mean level spacing is given by $\delta \sim L^{-\Delta}$, with $\Delta=1$, while in the complex case $\Delta=1/2$.
  \item Following the previous examples, at any $b$ the derivative $df(b)/db$ cannot be larger than $1$ for the real case and $2$ for the complex one.
\end{itemize}

\section{Phase diagram for other values of $d$}
\label{app:appB}
\begin{figure}
    \centering
    \includegraphics[width=0.45\columnwidth]{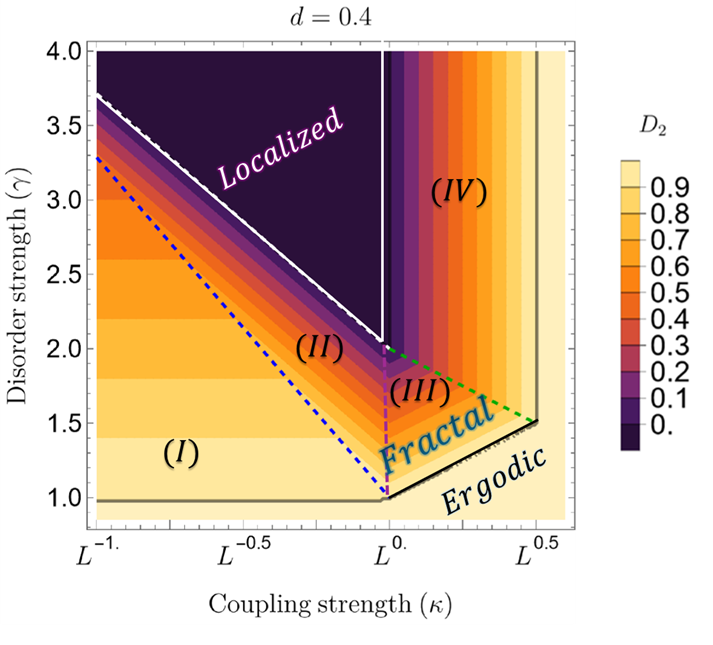}
    \includegraphics[width=0.45\columnwidth]{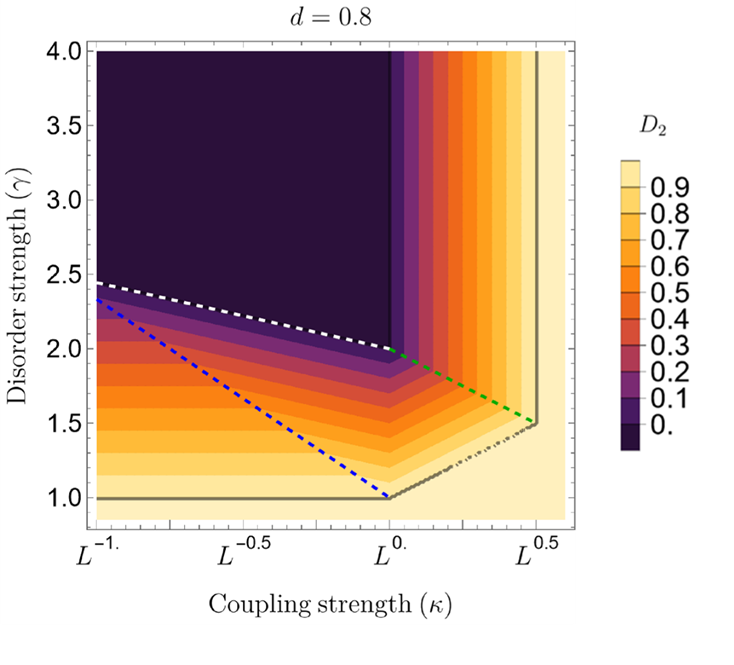}
    \caption{Plots showing variation of $D_2$ vs disorder strength represented by $\gamma$ and nearest-neighbor coupling strength $\kappa$ for fractal disorder with fractal dimensions left: $d=0.4$ and right: $d=0.8$.}
    \label{fig:comparisonschematic}
\end{figure}

In Fig.~\ref{fig:comparisonschematic} we show a comparison between how the reentrant phase diagram varies upon variation of the dimension of the fractal diagonal disorder $d$. We plot the scenario for (a)~a low, $d=0.4$, and (b)~high, $d=0.8$, fractal dimension of the on-site disorder. Note that $d>1$ gives same results as $d=1$. We immediately notice that as the $d$ increases the localized phase increases and the re-entrant region shrinks on the phase diagram. This is completely consistent with expectations as to put it briefly, the reentrance happens due to the sensitivity of the local-in-energy Fermi's golden rule broadening $\Gamma_d$ to the fractal on-site disorder and the interplay of the former $\Gamma_d$ to the local-in-space hopping term $\kappa$, fully insensitive to any fine spectral structure.
However since $d=1$ has the same characteristic behavior as the standard RP model, no additional fine spectral structure emerges at $d=1$. Thus, it should come as no surprise that the region shrinks as increase $d$ towards that value.

\end{document}